\documentclass[letterpaper,twocolumn,10pt]{article}
\usepackage{usenix,epsfig,endnotes}
\usepackage{pgfplots}
\usepackage{amsmath,amssymb}
\usepackage[capitalize]{cleveref}

\begin{document}

\date{}

\title{\Large \bf Performance evaluation of the Mojette erasure code for fault-tolerant distributed hot data storage}

\author{
{\rm Dimitri Pertin}\\
Universit\'{e} de Nantes\\
IRCCyN UMR 6597\\
Rozo Systems\\
\and
{\rm Didier F\'{e}ron}\\
Rozo Systems\\
\and
{\rm Alexandre Van Kempen}\\
Universit\'{e} de Nantes\\
IRCCyN UMR 6597\\
\and
{\rm Beno\^{i}t Parrein}\\
Universit\'{e} de Nantes\\
IRCCyN UMR 6597\\
} 

\maketitle

\thispagestyle{empty}

\subsection*{Abstract}
Packet erasure codes are today a real alternative to replication in fault
tolerant distributed storage systems. In this paper, we propose the Mojette
erasure code based on the Mojette transform, a formerly tomographic tool. The
performance of coding and decoding are compared to the Reed-Solomon code
implementations of the two open-source reference libraries namely ISA-L and
Jerasure 2.0. Results clearly show  better performances for our discrete
geometric code compared to the classical algebraic approaches. A gain factor up
to $2$ is measured in comparison with the ISA-L Intel . Those very good
performances allow to deploy Mojette erasure code for hot data distributed
storage and I/O intensive applications.

\section{Introduction}

Storage systems rely on redundancy to face ineluctable data unavailability and component failures. For its simplicity, data replication is the de facto
standard to provide redundancy. For instance, three-way replication is the
storage policy adopted by major file systems such as
Hadoop Distributed File System~\cite{shvachko2010msst} and Google File
System~\cite{ghemawat2003sosp}. While being straightforward to implement, plain
replication typically incurs high storage overheads. It has now been
acknowledged that erasure codes can significantly reduce the amount of
redundancy compared to replication while offering the same data
protection~\cite{weatherspoon2001iptps}.

However, these storage savings come at a price in terms of additional
complexity, as data must be encoded during write operations, and decoded during
read operations. Very efficient coding operations are thus keys to maintain
transparent operations for I/O intensive applications.
Since data replication has higher storage costs but performs faster than
erasure codes, storage systems tend to differentiate between \emph{cold} data
(i.e. not frequently accessed, such as in long term storage) and \emph{hot}
data, typically data that is frequently accessed.
In practice, plain replication is used for I/O intensive applications due to
fast data accesses while erasure codes are limited to long-term storage because
of their extra complexity.

\textit{Reed-Solomon} (RS) are the most popular codes as they provide
deterministic general-purpose codes without limit on the parity level. They are mostly implemented in their systematic form, meaning that the information data is a part of the encoded data. In addition, they are known to be Maximum Distance Separable (MDS) thus providing the optimal reliability for a given storage overhead. 
The former definition of RS codes are based on Vandermonde matrices
and expensive Galois field operations. Implementing such codes in an efficient
manner is therefore challenging. One of the best known implementation is
provided by Jerasure~\cite{plank2014jerasure}, an open-source library that
relies on Cauchy generating bit-matrices to only perform XOR operations, thus
avoiding the costly multiplications. Recently, Intel released ISA-L, a
performance-oriented open-source library~\cite{isa-l} that implements
RS codes leveraging SIMD instructions. To the best of our knowledge, these two
are the most efficient implementations publicly available.

In this paper, we propose to use the \textit{Mojette}
transform~\cite{guedon2005springer}, a formerly tomographic tool, to implement
a high-performance erasure code. 
The Mojette transform is a discrete and
exact version of the Radon transform and relies on discrete geometry, contrary to the classic algebraic code definition. By nature, the Mojette transform provides a non-systematic erasure code.
The geometric
approach coupled with an optimized implementation help to perform very fast encoding and decoding
operations, handling I/O intensive applications such as
virtualization or databases, that access small blocks of data ($4$~KB or
$8$~KB) in a random pattern~\cite{pertin2014closer}. Those block sizes fit the
general-purpose file systems requirements such as \textit{ext4} or
\textit{Btrfs}.

\section{The Mojette Erasure Code}
\label{sec:mojette}
 \vspace{-0.3cm}

This section presents how the Mojette transform is used to encode data,  
the uniqueness conditions of the reconstruction solution and its
inverse algorithm enabling the decoding. 
Finally, the end of this section details how the Mojette transform
is used as an erasure code in practical systems.

    \subsection{Forward Mojette Transform}

\begin{figure}
  \centering
  \includegraphics[width=0.9\columnwidth]{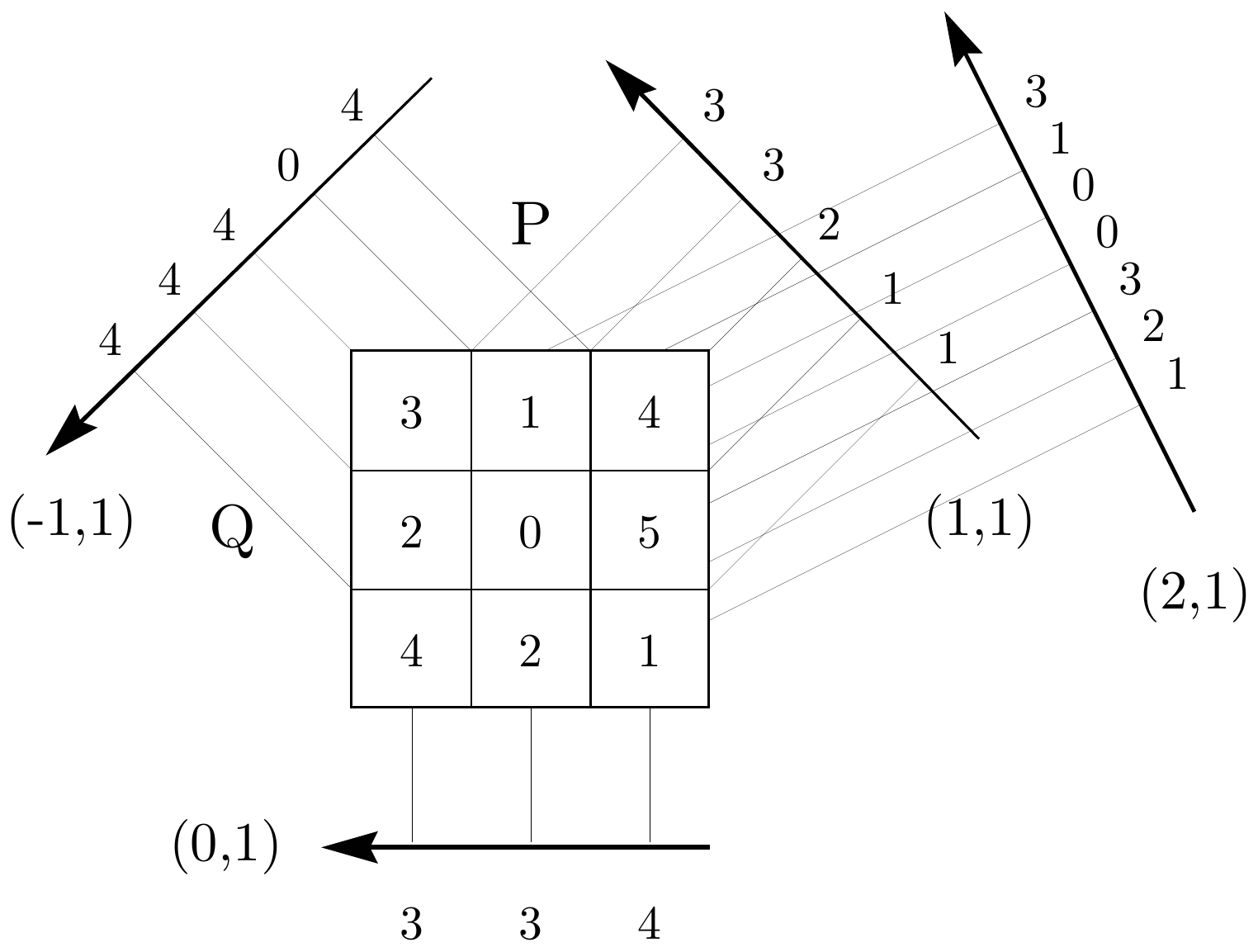}
  \caption{Mojette transform of a $3 \times 3$ image for directions $(p,q)$ in
      the set $\left\{(-1,1), (0,1), (1,1), (2,1)\right\}$. Addition is done
  here modulo $6$.}
  \label{fig:mojette}
\end{figure}

The forward Mojette transform is a linear operation that computes a set of 1D projections at
different angles, from a discrete image $f:(k,l)\mapsto\mathbb N$, composed
of $P \times Q$ \emph{pixels}. A projection direction is defined by a couple
of co-prime integers $(p,q)$. Projections are vectors of variable sizes whose
elements are called \emph{bins}. A bin in the Mojette transform
of $f$ is characterised by its position $b$ in
the projection which corresponds to a discrete line of equation $ b = -kq + lp$.
Its value is the sum of the centered pixels along the line:
\begin{equation}
    (M_{(p,q)}f)(b) = \sum^{P-1}_{k=0} \sum^{Q-1}_{l=0} f(k,l) [b=-kq+lp],
    \label{eqn:mojette}
\end{equation}
where, $[\cdot]$ is the Iverson bracket ($[P]=1$ whenever $P$ is true, $0$ otherwise).
The number of bins $B$ of a projection depends on the projection direction
$(p,q)$ and the lattice size $P \times Q$:
\begin{equation}
    B(p,q,P,Q) = \left|p\right|(Q-1) + \left|q\right|(P-1) + 1.
    \label{eqn:projection_size}
\end{equation}

Figure~\ref{fig:mojette} gives an example of the forward Mojette transform for a $3
\times 3$ integer image. The process transforms the 2D image into a set
of $I=4$ projections along the directions of the following set:
$\left\{(-1,1), (0,1), (1,1), (2,1)\right\}$. For the sake of this example,
addition is arbitrarily done modulo-$6$ (but any addition works). The
complexity $\mathcal{O}(PQI)$ is linear with the number of projections and the
number of grid elements. Note that some border bins are the exact copy of some
pixels. This remark will help to understand how starts the inverse transform
algorithm.
%

    \subsection{Inverse Mojette Transform}
    \label{subsec:inverse}

In this section, we first expose the reconstruction criterion on the projection set which yields to a unique reconstructed image. Then, we detail 
how is implemented the reconstruction algorithm.

\paragraph{Reconstruction Criterion}
\label{subsubsec:reconstruction}
Katz has shown that for a $P \times Q$ lattice, the reconstruction is possible
given a projection set $ S_I $ if one of the following criterion is
verified~\cite{katz1978springer}:
\begin{equation}
    P \leq \sum^{I-1}_{i=0} |p_i| \text{ or } Q \leq \sum^{I-1}_{i=0} |q_i|,
    \label{eqn:katz}
\end{equation}
where $I$ is the number of projections involved in the reconstruction process. 


In the example of the Figure~\ref{fig:mojette}, we see that each subset of $3$ projections
$\left\{(p_{0},q_{0}),\dots,(p_{2},q_{2})\right\}$ is such that
$\sum_{i=0}^2~|q_{i}|=3$. Thus, the $4$ projections in Figure~\ref{fig:mojette} depicts a redundant
representation of the image, where any $3$ projections among these $4$ can be used for reconstruction.


\paragraph{Inverse Mojette Algorithm}
\label{subsubsec:inverse_algorithm}
\vspace{-0.3cm}


The reconstruction algorithm aims at finding a reconstructible bin and to
write its value in the image by back-projection. Bins are reconstructible
when they result from a unique pixel of the image. Once a bin is reconstructed,
its contribution is subtracted from all the projections involved in the
reconstruction, thus paving the way to reconstruct further bins. As the forward
algorithm, the Mojette inverse is linear with the number of projections $I$ and
the number of elements $P \times Q$ in the grid.

Observing that the reconstruction propagates from the image corners to its
center, Normand et al.~\cite{normand2006dgci} showed that given an image domain and a projection set,
a dependency graph between the image pixels can be found.
Within this graph, considering that a single projection is dedicated
to the reconstruction of a single line of the image, a reconstruction path can
be pre-determined. We refer the interested reader to~\cite{normand2006dgci} due to lack of space. 


    \subsection{Properties of the Mojette Erasure Code}
    \label{subsec:mojette_erasure_code}

The Mojette erasure code extends the application of the Mojette transform, originally designed for images, to any type of data.
As the Mojette transform creates a redundant representation, it appears to be an appealing candidate to provide failure tolerance in storage systems. In classic coding theory words, we consider the $k$ lines of the Mojette array as the input data packets, and we compute $n$ projections as the set of encoded packets. The Mojette erasure code is therefore non-systematic here. Note that a systematic version of the Mojette is currently under development. Since the size of projection varies with the parameters $(p,q)$ we
consider for each projection that $q_i=1$ to limit the bin overhead (as proposed
in~\cite{parrein2001dcc}). Then the Katz criterion proves that if we get any $k$ out of the $n$ projections, it is possible to exactly reconstruct the array. This way, the storage system is able to face the unavailability of up to any $n-k$ storage nodes. 

In practice, we can observe that some bins are never used during reconstruction whatever the projection set used for the process~\cite{verbert}. Removing these bins from the
encoding process, particularly when $p$ increases, significantly limits the
projection size variation and therefore yields to a negligible storage overhead
relative to the MDS case.

\vspace{-0.3cm}
\section{Erasure Code Micro-benchmark}
\label{sec:performance}

In this section, we evaluate the performance of our new erasure code compared
to the Jerasure and ISA-L libraries. Firstly, we describe our Mojette
implementation and then present our two competitors. Secondly, we detail the
experiment setup to finally depict the results and analysis in the last
section.
\vspace{-0.3cm}
\paragraph{Mojette}
We implemented a non-systematic version of the Mojette erasure code in C.
In practice, pixel size should fit a computer word to improve
performance based on XOR operations. Since x86 architectures provide Streaming
SIMD Extensions (SSE) instruction set, pixel and bin sizes are set to 128 bits
to benefit from high-performance XOR computations.
The Mojette encoding requires at most $k-1$ XORs per computed bin (and zero XOR
for bins at projection edges). Similarly for decoding, at most $k-1$ XORs are
required per reconstructible pixel. The progressive reconstruction from left to
right of connected pixels (as proposed in~\cite{normand2006dgci}), coupled with
a drastic reduction of updates, guarantees spatial memory locality, thus
high-performance computation.
\vspace{-0.3cm}
\paragraph{Jerasure}
The first competitor is the systematic Vandermonde implementation of RS
codes from the open-source Jerasure 2.0 library~\cite{plank2014jerasure}. We
choose their Vandermonde implementation since it performs better than their
Cauchy-based implementation for such small packets size.
%
%
The Galois field size is set to $w=8$ to fit our erasure code configuration $(n,k)$.

\paragraph{Intel ISA-L}
The second competitor is the RS implementation provided in Intel
ISA-L open-source library~\cite{isa-l}. It is one of the fastest systematic
erasure code implementation since it makes intensive usage of the x86
architecture features such as \emph{xmm} registers and SSE instructions.\\



\vspace{-0.6cm}
\subsection{Experiment Setup}


We conducted all experiments on small data blocks of \textit{BlockSize} equals to $4$~KB and $8$~KB (that fit block-based file-system like \textit{ext4}). 
For encoding, we consider a single data block filled with random data. For decoding, we record the performance as we increase the number of erasures
up to the failure tolerance. 
With systematic codes (implementations of ISA-L and Jerasure), erasures only concern data packets (no decoding otherwise). 


Two erasure code configurations are considered for the benchmark: $(n,k)$
equals $(6,4)$ and $(12,8)$, preventing from $2$ and $4$ failures respectively.
All the computations are performed in memory, with no disk I/O operation.
Furthermore, we do not take into account pre-computations such as the matrix
inversion or deterministic reconstruction path respectively for the RS and
Mojette implementations.
Since we measure optimized encoding and decoding functions that are mostly
computation-bounded, with the data  entirely located in L1 and L2 cache, we use the \emph{RDTSC} instruction that returns the \emph{time stamp counter} (TSC) which is incremented on every CPU clock cycle~\cite{intel1997rdtsc}. For all tested implementations, the standard deviation is too negligible to be represented (less than 1\%).

All the experiments are done on a single processor running Linux $3.2$ and Debian
Wheezy over an x86-64 architecture. It embeds a $1.80$~GHz Intel Xeon
processor, with $16$~GB of RAM and cache sizes of $32$, $256$ and $10240$~KB
for respectively L1, L2 and L3 cache levels.

\subsection{Results}

We now present the results of encoding and decoding throughput for various
\textit{BlockSize} and code parameters. 
For the sake of comparison, we plot the optimal performance recorded by the 
\textbf{memcpy()}. More precisely, the optimal encoding is given by the \textbf{memcpy()} of $n$ packets of
$\frac{\textit{BlockSize}}{k}$ bytes while the optimal decoding is the \textbf{memcpy()} of only $k$ packets among the $n$ encoded.
Once again, note that the Mojette is implemented as a non-systematic code, thus increasing the overall computation compared to the two other systematic codes.


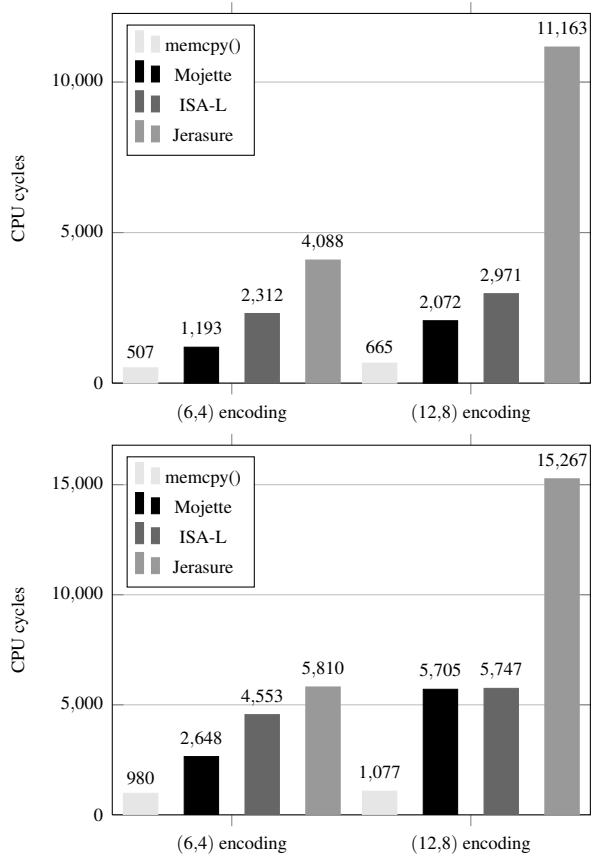
\begin{figure}[t!]
  \centering
  \pgfplotsset{width=\linewidth, height=6.5cm}
  \begin{tikzpicture}
 \tikzstyle{every node}=[font=\scriptsize]
\begin{axis}[
    ybar=10pt,              
    enlarge x limits=0.5,   
    legend pos=north west,
    legend style={legend columns=1},
    ylabel={CPU cycles},   
    symbolic x coords={1,2},
    bar width=13pt,
    ymin=0,
    xtick=data,%
    xticklabels={$(6{,}4)$ encoding,$(12{,}8)$ encoding},
    error bars/.cd,
    yminorgrids=true,
    ymajorgrids=true,
    nodes near coords,
    scaled y ticks=base 10:0,   
    cycle list = {black!10,black,black!60,black!40},
]

\addplot+[fill, text=black, error bars/.cd, y dir=both, y explicit]
 coordinates {
    (1,507)
    (2,665)
};

\addplot+[fill, text=black, error bars/.cd, y dir=both, y explicit]
 coordinates {
    (1,1193)
    (2,2072)
};
\addplot+[fill, text=black, error bars/.cd, y dir=both, y explicit]
coordinates {
    (1,2312)
    (2,2971)
};
\addplot+[fill, text=black, error bars/.cd, y dir=both, y explicit]
coordinates {
    (1,4088)
    (2,11163)
};

\legend{memcpy(), Mojette, ISA-L, Jerasure, }

\end{axis}
\end{tikzpicture}

%
  \centering
  \pgfplotsset{width=\linewidth}
  \begin{tikzpicture}
 \tikzstyle{every node}=[font=\scriptsize]
\begin{axis}[
    ybar=10pt,              
    enlarge x limits=0.5,   
    legend pos=north west,
    legend style={legend columns=1},
    ylabel={CPU cycles},
    symbolic x coords={1,2},
    bar width=13pt,
    ymin=0,
    xtick=data,%
    xticklabels={$(6{,}4)$ encoding,$(12{,}8)$ encoding},
    error bars/.cd,
    yminorgrids=true,
    ymajorgrids=true,
    nodes near coords,
    scaled y ticks=base 10:0,   
    cycle list = {black!10,black,black!60,black!40}
]

\addplot+[fill, text=black, error bars/.cd, y dir=both, y explicit]
 coordinates {
    (1,980)
    (2,1077)
};

\addplot+[fill, text=black, error bars/.cd, y dir=both, y explicit]
 coordinates {
    (1,2648)
    (2,5705)
};
\addplot+[fill, text=black, error bars/.cd, y dir=both, y explicit]
coordinates {
    (1,4553)
    (2,5747)
};
\addplot+[fill, text=black, error bars/.cd, y dir=both, y explicit]
coordinates {
    (1,5810)
    (2,15267)
};

\legend{memcpy(), Mojette, ISA-L, Jerasure}

\end{axis}
\end{tikzpicture}
  \caption{Encoding performance for an input data block of $4$~KB
  (\textbf{top}) and $8$~KB (\textbf{bottom}) depending on the code parameters $(n=6,k=4)$ or $(n=12,k=8)$. }
   \label{fig:bench_enc4k}
\end{figure}

\paragraph{Encoding}
\vspace{-0.3cm}
Figure \ref{fig:bench_enc4k} shows the encoding
performance recorded for $4$~KB (top) and $8$~KB (bottom) data blocks for the
$(6,4)$ and $(12,8)$ codes. The first observation is that the Mojette erasure
code outperforms the two other implementations in every tested settings. For
example, to encode a $4$~KB block with a $(6,4)$ code,
the Mojette implementation divides the number of CPU cycles by a factor of $1.94$ and $3.42$ when respectively compared to ISA-L and Jerasure. While the Mojette implementation still provides the closest performance from the optimal value of the \textbf{memcpy()},
the improvements are mitigated for the code $(12,8)$, and especially versus
ISA-L for a block of $8$~KB. This is mainly due to our non-systematic design. 
Indeed, to encode a  $8$~KB block with a $(12,8)$ code, the encoder splits $8$~KB into $8$ packets of one kilobyte and produces $4$ encoded packets for systematic codes, while non-systematic codes have to compute $12$ encoded packets thus performing $3$ times more computations.
Finally, we notice that, as expected, the CPU cycles number linearly increases with the \textit{BlockSize}, as well as with the number of blocks to be encoded thus experimentally confirming the linear complexity of the Mojette transform. 


\paragraph{Decoding}
\vspace{-0.3cm}
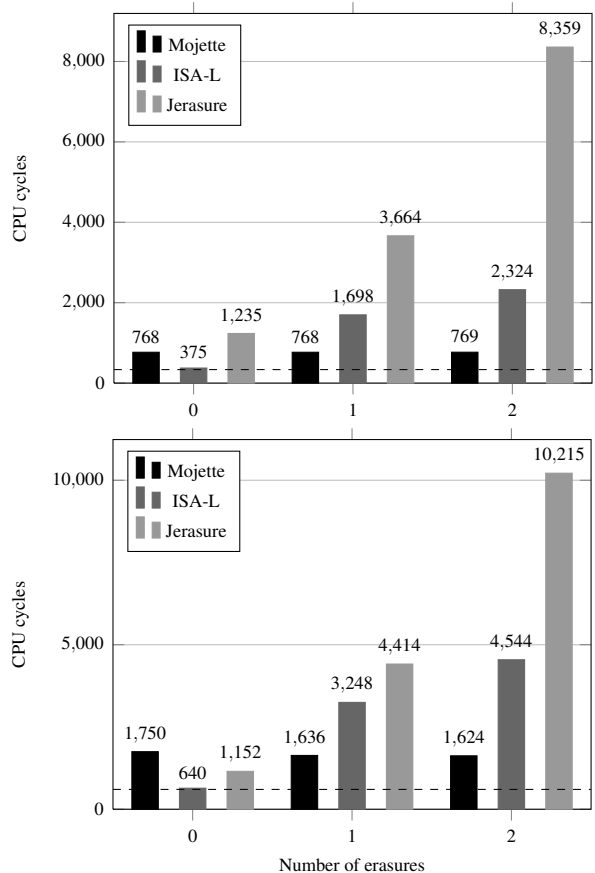
\begin{figure}[t!]
  \centering
  \pgfplotsset{width=\linewidth, height=6.5cm}
  \begin{tikzpicture}
 \tikzstyle{every node}=[font=\scriptsize]
\begin{axis}[
    ybar=8pt,              
    enlarge x limits=0.25,   
    legend pos=north west,
    legend style={legend columns=1},
    ylabel={CPU cycles},
    symbolic x coords={1,2,3},
    bar width=10pt,
    ymin=0,
    xtick={1,2,3},%
    xticklabels={0,1,2},
    error bars/.cd,
    yminorgrids=true,
    ymajorgrids=true,
    nodes near coords,
    cycle list = {black,black!60,black!40,black!10},
]

\addplot+[fill, text=black, error bars/.cd, y dir=both, y explicit]
 coordinates {
    (1,768)
    (2,768)
    (3,769)
};
\addplot+[fill, text=black, error bars/.cd, y dir=both, y explicit]
coordinates {
    (1,375)
    (2,1698)
    (3,2324)
};
\addplot+[fill, text=black, error bars/.cd, y dir=both, y explicit]
coordinates {
    (1,1235)
    (2,3664)
    (3,8359)
};

\draw [black,dashed] ({rel axis cs:0,0}|-{axis cs:2,336}) -- ({rel axis cs:1,0}|-{axis cs:2,336}) node [pos=0.33, above] {};

\legend{Mojette, ISA-L, Jerasure}

\end{axis}
\end{tikzpicture}

%
  \centering
  \pgfplotsset{width=\linewidth}
  \begin{tikzpicture}
 \tikzstyle{every node}=[font=\scriptsize]
\begin{axis}[
    ybar=8pt,              
    enlarge x limits=0.25,   
    legend pos=north west,
    legend style={legend columns=1},
    ylabel={CPU cycles},
    xlabel={Number of erasures},
    symbolic x coords={1,2,3},
    bar width=10pt,
    ymin=0,
    xtick={1,2,3},%
    xticklabels={0,1,2},
    error bars/.cd,
    yminorgrids=true,
    ymajorgrids=true,
    nodes near coords,
    scaled ticks=base 10:0,     
    cycle list = {black,black!60,black!40,black!10},
]

\addplot+[fill, text=black, error bars/.cd, y dir=both, y explicit]
 coordinates {
    (1,1750)
    (2,1636)
    (3,1624)
};
\addplot+[fill, text=black, error bars/.cd, y dir=both, y explicit]
coordinates {
    (1,640)
    (2,3248)
    (3,4544)
};
\addplot+[fill, text=black, error bars/.cd, y dir=both, y explicit]
coordinates {
    (1,1152)
    (2,4414)
    (3,10215)
};

\draw [black,dashed] ({rel axis cs:0,0}|-{axis cs:2,603}) -- ({rel axis cs:1,0}|-{axis cs:2,603}) node [pos=0.33, above] {};

\legend{Mojette, ISA-L, Jerasure}

\end{axis}
\end{tikzpicture}

  \caption{Decoding performance of a $(n=6,k=4)$ code for an input data block
  of $4$~KB (\textbf{top}) and $8$~KB (\textbf{bottom}) depending on the number
  of failures. The dashed line depicts the optimal value of the
  \textbf{memcpy()} respectively measured at 336 ($4$~KB) and 603 ($8$~KB).}
   \label{fig:bench_dec4k1}
\end{figure}

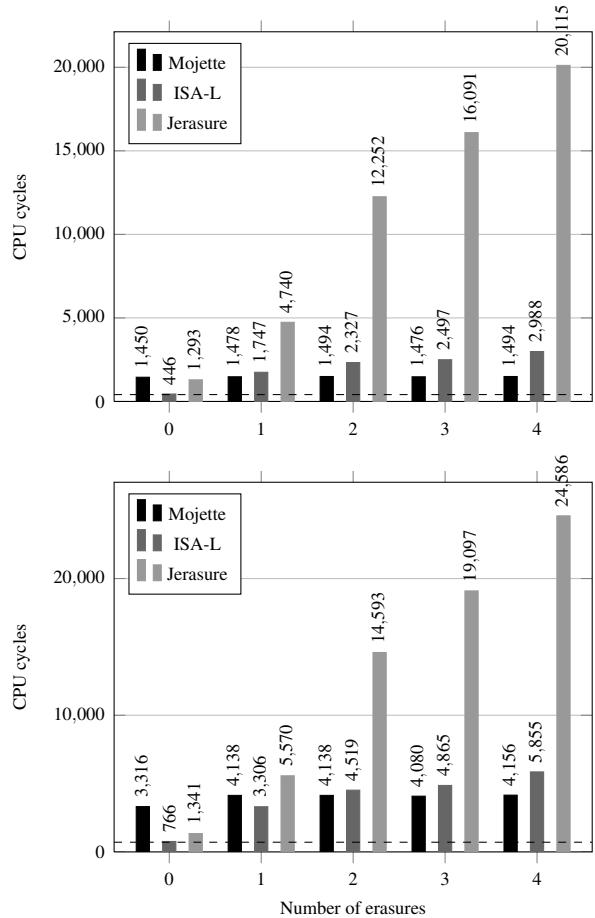
\begin{figure}[t!]
  \centering
  \pgfplotsset{width=\linewidth, height=6.5cm}
    \begin{tikzpicture}
 \tikzstyle{every node}=[font=\scriptsize]
\begin{axis}[
    ybar=5pt,              
    enlarge x limits=0.15,   
    legend pos=north west,
    legend style={legend columns=1},
    ylabel={CPU cycles},
    symbolic x coords={1,2,3,4,5},
    bar width=5pt,
    ymin=0,
    xtick={1,2,3,4,5},%
    xticklabels={0,1,2,3,4},
    error bars/.cd,
    yminorgrids=true,
    ymajorgrids=true,
    nodes near coords,
    every node near coord/.append style={rotate=90, anchor=west},
    scaled ticks=base 10:0,     
    cycle list = {black,black!60,black!40,black!10},
]

\addplot+[fill, text=black, error bars/.cd, y dir=both, y explicit]
 coordinates {
    (1,1450)
    (2,1478)
    (3,1494)
    (4,1476)
    (5,1494)
};
\addplot+[fill, text=black, error bars/.cd, y dir=both, y explicit]
coordinates {
    (1,446)
    (2,1747)
    (3,2327)
    (4,2497)
    (5,2988)
};
\addplot+[fill, text=black, error bars/.cd, y dir=both, y explicit]
coordinates {
    (1,1293)
    (2,4740)
    (3,12252)
    (4,16091)
    (5,20115)
};

\draw [black,dashed] ({rel axis cs:0,0}|-{axis cs:2,411}) -- ({rel axis cs:1,0}|-{axis cs:2,411}) node [pos=0.33, above] {};

\legend{Mojette, ISA-L, Jerasure}

\end{axis}
\end{tikzpicture}

%
%
  \centering
  \pgfplotsset{width=\linewidth}
  \begin{tikzpicture}
 \tikzstyle{every node}=[font=\scriptsize]
\begin{axis}[
    ybar=5pt,              
    enlarge x limits=0.15,   
    legend pos=north west,
    legend style={legend columns=1},
    ylabel={CPU cycles},
    xlabel={Number of erasures},
    symbolic x coords={1,2,3,4,5},
    bar width=5pt,
    ymin=0,
    xtick={1,2,3,4,5},%
    xticklabels={0,1,2,3,4},
    error bars/.cd,
    yminorgrids=true,
    ymajorgrids=true,
    nodes near coords,
    every node near coord/.append style={rotate=90, anchor=west},
    scaled ticks=base 10:0,     
    cycle list = {black,black!60,black!40,black!10},
]

\addplot+[fill, text=black, error bars/.cd, y dir=both, y explicit]
 coordinates {
    (1,3316)
    (2,4138)
    (3,4138)
    (4,4080)
    (5,4156)
};
\addplot+[fill, text=black, error bars/.cd, y dir=both, y explicit]
coordinates {
    (1,766)
    (2,3306)
    (3,4519)
    (4,4865)
    (5,5855)
};
\addplot+[fill, text=black, error bars/.cd, y dir=both, y explicit]
coordinates {
    (1,1341)
    (2,5570)
    (3,14593)
    (4,19097)
    (5,24586)
};

\draw [black,dashed] ({rel axis cs:0,0}|-{axis cs:2,711}) -- ({rel axis cs:1,0}|-{axis cs:2,711}) node [pos=0.33, above] {};

\legend{Mojette, ISA-L, Jerasure}

\end{axis}
\end{tikzpicture}
  \caption{Decoding performance of a $(n=12,k=8)$ code for an input data block
  of $4$~KB (\textbf{top}) and $8$~KB (\textbf{bottom}) depending on the number
  of failures. The dashed line depicts the optimal value of the
  \textbf{memcpy()} respectively measured at 411 ($4$~KB) and 711 ($8$~KB).}
   \label{fig:bench_dec8k1}
\end{figure}

We respectively plot in Figure~\ref{fig:bench_dec4k1,fig:bench_dec8k1}
the number of CPU cycles required to decode the data for the same codes as before, depending on the number of failures. We still emphasize here the differences between systematic and non-systematic implementations. Since RS codes are systematic, \textbf{when no failure occurs}, they should achieve optimal performance (equivalent to a \textbf{memcpy()}) as the decoding process boils down to the copy of $k$ data blocks in memory. For example, we see that ISA-L delivers the optimal performance for every $0$-erasure settings. 
Note that our ongoing implementation of the Mojette in systematic-form would also provide the same results.

We now focus on the results in the presence of failures, when decoding operations are therefore involved (i.e. we do not just retrieve the data packets in memory).
Results for the code $(n=6,k=4)$ on a $4$~KB block, depicted on top of the Figure~\ref{fig:bench_dec4k1} show that the number of CPU cycles is divided
by a factor of $2.2$ and $4.8$ when respectively compared to ISA-L and Jerasure for a single failure.
These factors are even higher when two erasures occurred.
In fact, due to its non-systematic form, the set of projections used has no influence on the decoding performances of the Mojette. On the contrary, the performances of Jerasure and ISA-L progressively decrease with the number of erasures.
Although this performance gap is reduced for the code $(n=12,k=8)$, especially versus ISA-L, results presented in Figure~\ref{fig:bench_dec8k1} still confirm the above observations.



%
%



\vspace{-0.3cm}
\section{Conclusion}
\label{sec:conclusion}
\vspace{-0.3cm}
Erasure codes are well known to incur a high computational penalty due to their
inherent coding operations, thus preventing them from being deployed  in I/O
intensive applications. In this paper, we advocated that the \textit{Mojette
transform} is a particularly suitable tool to design high-performance erasure
code. We implemented and evaluated our new erasure code compared to the
best-known implementations, namely ISA-L and Jerasure. Results show that this
paradigm shift towards a geometric approach enables the \textit{Mojette}-based
implementation to significantly improve the throughput of coding and decoding
operations. As non-systematic, the proposed code can still bring better throughputs in a foreseeable future.
A Mojette erasure code implementation is currently deployed in an open-source project RozoFS~\cite{rozofs}. We believe that this new code paves the way to the use of erasure codes in I/O intensive applications.

\vspace{-0.3cm}
\section{Acknowledgements}
\label{sec:acknowledgments}
\vspace{-0.3cm}
This material is based upon work supported by the Agence Nationale de la
Recherche (ANR) through the project FEC4Cloud (ANR-12-EMMA-0031-01).

\vspace{-0.2cm}
{\footnotesize \bibliographystyle{acm}
\bibliography{pertin2015hotstorage}}

\end{document}